\documentclass[twocolumn,preprintnumbers,pre,longbibliography]{revtex4-1}

\usepackage[caption=false]{subfig}
\usepackage[countmax]{subfloat}
\usepackage{graphicx}
\usepackage{amssymb, amsmath,amssymb,amsfonts,mathrsfs,amsopn}
\usepackage{amsthm,mathrsfs,amsopn}
\usepackage{dcolumn}
\usepackage{bm}
\usepackage{color}
\usepackage[utf8]{inputenc}
\usepackage{amssymb, amsmath,amssymb,amsfonts}
\usepackage{bm}
\usepackage[table]{xcolor} 
\usepackage{floatrow}
\usepackage{algorithmicx}
\usepackage{algorithm}
\usepackage{algpseudocode}
\usepackage{mathtools}
\usepackage{fancyhdr} 
\begin{document}

\title{{\Large Structure and {dynamical behaviour} of non-normal networks}}

\author{Malbor Asllani$^{1,2}$\footnote{malbor.asllani@unamur.be}, Renaud Lambiotte$^{1}$\footnote{renaud.lambiotte@maths.ox.ac.uk}, Timoteo Carletti$^2$}
\affiliation{$^1$Mathematical Institute, University of Oxford, Woodstock Rd, OX2 6GG Oxford, UK}
\affiliation{$^2$Department of Mathematics and naXys, Namur Institute for Complex Systems, University of Namur,\\
rempart de la Vierge 8, B 5000 Namur, Belgium}

\begin{abstract}
We analyse a collection of empirical networks in a wide spectrum of disciplines and show that strong non-normality is ubiquitous in network science. Dynamical processes evolving on non-normal networks exhibit a peculiar behaviour, as initial small disturbances may undergo a transient phase and be strongly amplified in linearly stable systems. Additionally, eigenvalues may become extremely sensible to noise, {and have a diminished physical meaning}. We identify structural properties of networks that are associated to non-normality and 
propose simple models to generate  networks with a tuneable level of non-normality. We also show the potential use of a variety of metrics capturing different aspects of non-normality, and propose their potential use in the context of the stability of complex ecosystems.
\end{abstract}

\maketitle

%\section{Introduction}
%\label{sec:intro}

Network science \cite{net1,net3,newman} has emerged, in the last 20 years, as an essential framework to model and understand complex systems in {a} variety of disciplines, including physics \cite{net1}, economy \cite{econet}, biology \cite{netgen} and sociology \cite{socialnet}. At its core, network science views a system as a set of nodes that may be connected directly by an edge or indirectly by a succession of edges, thereby forming paths of interactions. The bridge between network structure and dynamics is generally {unraveled} by defining a linear dynamical model on the nodes; take for instance a random walk process, as a simple model of diffusion, or the linearisation around a critical point of a non-linear dynamical system \cite{may,allesina,zoology,master}. In each case, the process is determined by a matrix, somehow related to the adjacency matrix of the underlying network. In addition, critical aspects of the system, such as its stability and characteristic time scales, are usually described by the properties of its spectrum \cite{van2010graph}. Central network concepts such as the spectral gap, spectral radius and Master Stability conditions all build on this interpretation. Relatedly, network spectra also appear in network algorithms, such as in community detection \cite{newman2006finding} or network comparison \cite{donnat2018tracking}. 

The characterisation of a linear system by its spectrum is canonical, but it is unreliable in situations when the linear operator is non-normal, {namely} its eigenvectors do not necessarily form an orthonormal basis and the transformation to eigenvector coordinates may involve a strong distortion of the phase space. Non-normality has a long tradition in linear algebra and dynamical systems, from early studies in hydrodynamics \cite{pseudospectra} to more recent works on the robustness of non-normal ecosystems \cite{ecoresilience} and in neuronal dynamics~\cite{Murphy2009,Hennequin2012}.  Yet, these results remain focused on limited areas of science and  a systematic study of the prevalence of non-normality in real-world networks, and its potential impact on dynamics, is still lacking. Here we call non-normal a network whose adjacency matrix $\textbf{A}$ is non-normal~\cite{nonnormal}. By definition, $\textbf{A}$ is non-normal if it verifies $\textbf{A} \textbf{A}^{T} \neq \textbf{A}^{T} \textbf{A}$.  It is thus clear that $\textbf{A}$ needs to be asymmetric to be non-normal, or equivalently the network needs to be directed to be non-normal, but, as we will discuss more in detail later, asymmetry is not sufficient and certain types of network architectures are {necessary to determine} a strong non-normality. It is also important to note that {given a non-normal network} other standard matrices, such as its Laplacian $\textbf{L}$, are also non-normal. {Non-normality can hence be quantified using a standard spectral measure borrowed from matrix theory, such as the Henrici's departure from normality~\cite{spectra}, $d_F(\mathbf{M})=\sqrt{||\mathbf{M}||_F^2-\sum_{i=1}^n|\lambda_i|^2}$, where $||\cdot||_F$ is the Frobenius norm and $\lambda_i$ the eigenvalues of the adjacency matrix. A zero value is associated to a symmetric network while the larger the values the stronger the non-normality.}

Before going further, let us briefly illustrate more in detail  the  influence of  non-normality on the prototypical example of a linear (linearised) dynamics on a non-normal network. Without loss of generality, we consider the model $\dot{\textbf{x}}= \textbf{Mx}$, where $\textbf{M}$ encodes the {linear} dynamics on the network, {through the dependence on $\textbf{L}$}, and forms a stable matrix, {namely the \textit{spectral abscissa} $\alpha(\textbf{M})=\max \Re\sigma(\textbf{M})$ is not positive, being $\sigma(\textbf{M})$ the spectrum of the matrix $\textbf{M}$.} 
In case of normal networks, the solution {of the linear system} would consist in a linear combination of exponentially relaxing modes, each with a characteristic time scale given by the inverse of the corresponding eigenvalue{, hence the spectral abscissa is responsible for the long time dynamics}. In situations when the network is non-normal, however, more complex patterns may emerge. Standard measures of non-normality of the matrix and their relation to dynamics are provided in Table~\ref{box:NN}.
Indeed, if  $\textbf{M}$ has a positive \textit{numerical abscissa}, {$\omega(\textbf{M})=\max\sigma(H(M))$, where $H(\textbf{M})=(\textbf{M}+\textbf{M}^T)/2$ is the Hermitian part of $\textbf{M}$}, the system can undergo a transient growth before asymptotically converging to zero, as measured by the norm of the state vector $\textbf{x}$ (see panel (a) in Table~\ref{box:NN}). This transient behaviour cannot be explained by the picture provided by {the spectrum of the matrix $\textbf{M}$}~\cite{spectra} and can have a strong impact once nonlinearities are at play.
In situations when the dynamics is obtained from the linearisation around a critical point, this initial growth may trigger nonlinear terms and take the system far away from the equilibrium, and thus radically reshape the dynamical behaviour of nonlinear systems \cite{nonnormal}, as shown in panel (b) in Table~\ref{box:NN}. 

\begin{table*}
\includegraphics[width=0.8\textwidth]{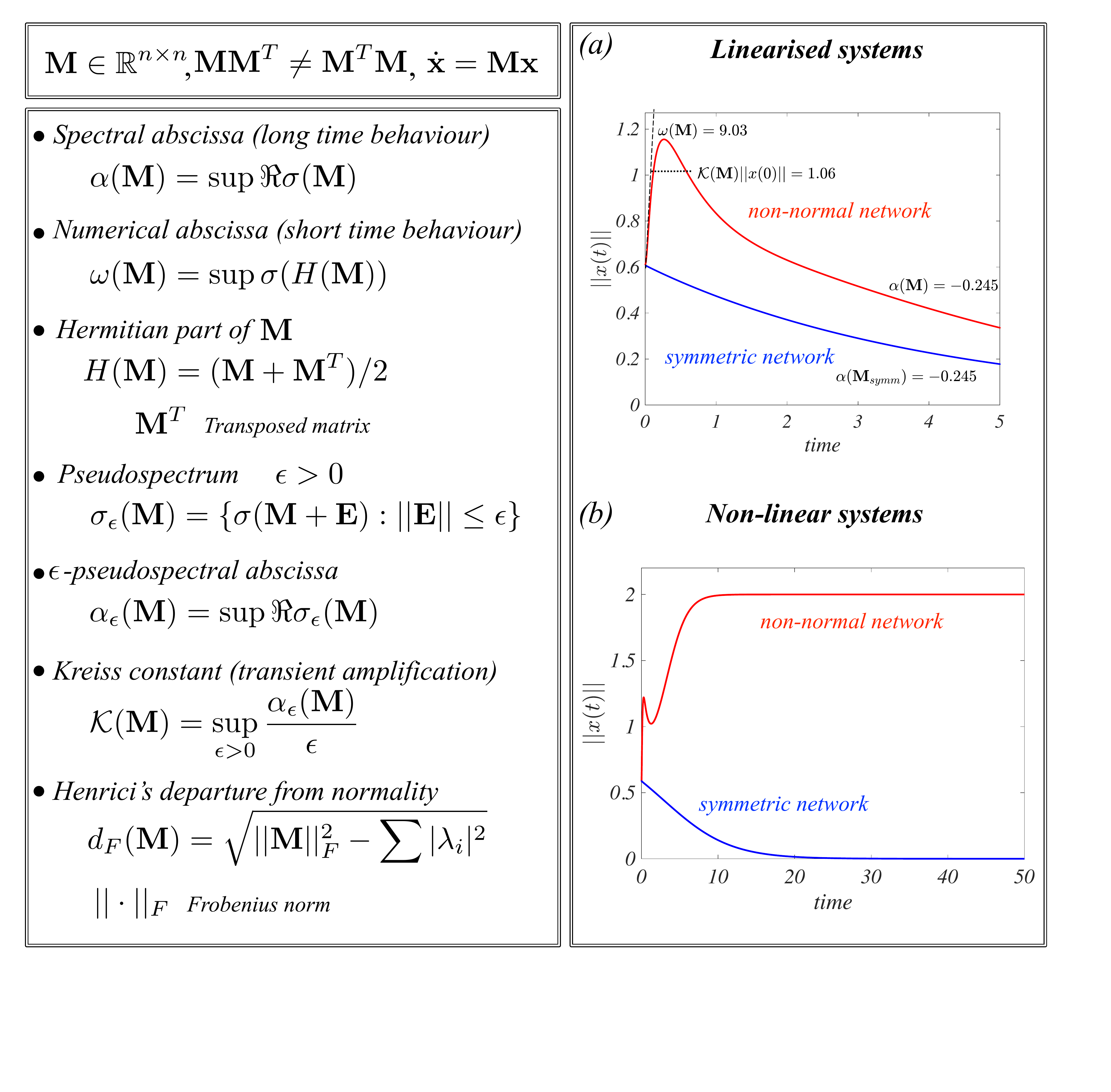}
\vspace*{-1cm}
\caption{\textbf{Summary of non-normal dynamics.} In the left panel, we summarise some of {the main concepts related to  non-normal dynamics. They are illustrated  through} the time evolution of the norm of the solution of the nonlinear bistable system $\dot{x}_i=f(x_i)+D (\mathbf{Lx})_i$ where $f(x)=x(a-x)(x-1)$ (panel b on the right) and its linearisation, $\dot{\mathbf{x}} = \mathbf{Mx}$, around the stable equilibrium $\mathbf{x}=0$ {(panel a on the right). Here $\mathbf{M}=a\mathbf{I}+D\mathbf{L}$, $a=-0.245$, $D=10$, $\mathbf{I}$ is the identity matrix and $\mathbf{L}$ is the Laplacian matrix of the underlying non-normal network~\cite{newman}}. Observe that the system is asymptotically stable, $\alpha(\mathbf{M})=a<0$. On both panels on the right, the red curves correspond to a non-normal Scale-Free network (nSF), while the blue ones to the symmetric version of the same network. Both systems are asymptotically stable and their {solutions} decrease to asymptotically reach the stable point (panel a) $\mathbf{x}=0$. However, one can appreciate the non-monotone convergence of the red curve due to an initial growth induced by a positive $\omega(\mathbf{M})$ and estimated by the Kreiss constant. This different behaviour has a striking consequence in the nonlinear model (panel b). In that case,  even if the equilibrium point remains stable, the initial amplification due to non-normality is sufficiently strong  to push the orbit towards another equilibrium {identified by an asymptotic positive amplitude $||x(t)||>0$ for $t\rightarrow \infty$}.}
\label{box:NN}
\end{table*}

\begin{table*}[t]
\begin{tabular}{l |c|c|c|c|c|c|c}
  \hline
\hspace{1em}{\small Network name} & {\small nodes} & {\small links} & {$\omega$} & {$\omega-\alpha$} & {$\alpha_{\epsilon}$} & {$\Delta$} & {$\hat{d}_F$}  \\
\hline
\textbf{Foodwebs} &   &   &   & &  &  &\\
\hspace{1em}{\small Cypress wetlands South Florida (wet)} & {\small $128$} & {\small $2016$}   & {\small $296.71$} & {\small $132.11$} & {\small $167.46$} & {\small $0.83$} & {\small $1.00$} \\  
\hspace{1em}{\small Cypress wetlands South Florida (dry) }& {\small $128$} & {\small $2137$}  & {\small $217.60$} & {\small $152.50$}& {\small $82.20$} & {\small $0.89$} & {\small $1.00$} \\
\hspace{1em}{\small Little Rock Lake (Wisconsin, US)} & {\small $183$} & {\small $2494$}   & {\small $21.69$} & {\small $14.69$}& {\small $10.02$} & {\small $0.95$} & {\small $0.93$} \\
\hline
\textbf{Biological} &   &   &   & &  &  &\\
\hspace{1em}{\small Transcriptional regulation network (\emph{E. coli})} & {\small $423$} & {\small $578$}   & {\small $5.11$} & {\small $4.11$}& {\small $2.52$} & {\small $0.81$} & {\small $0.93$}  \\
\hspace{1em}{\small Metabolic network (\emph{C. Elegans})} & {\small $453$} & {\small $4596$}   & {\small $13.44$} & {\small $12.44$}& {\small $6.89$} & {\small $0.98$} & {\small $1.00$} \\
\hspace{1em}{\small Pairwise proteins interaction ({\em Homo sapiens})} & {\small $2239$} & {\small $6452$}   & {\small $15.79$} & {\small $13.02$} & {\small $4.01$} & {\small $0.99$} & {\small $0.99$} \\
\hline
\textbf{Transport} &   &   &   & &  &  \\
\hspace{1em}{\small US airport 2010} & {\small $1574$} & {\small $28236$}   & {\small $1.19\, 10^7$} & {\small $79.30$}& {\small $1.19\, 10^7$} & {\small $0.01$} & {\small $1.00$}\\
\hspace{1em}{\small Road transportation network (Rome)} & {\small $3353$} & {\small $8870$}   & {\small $2.40\, 10^4$} & {\small $120.05$} & {\small $2.39\, 10^4$} & {\small $0.08$} & {\small $0.28$} \\
\hspace{1em}{\small Road transportation network (Chicago)} & {\small $12982$} & {\small $39018$}   & {\small $4.23$} & {\small $4.29\, 10^{-4}$}& {\small $4.54$} & {\small $0.04$} & {\small $0.19$} \\
\hline
\textbf{Communication} &   &   &  & &  & &\\
\hspace{1em}{\small e-mails network DNC} & {\small $2029$} & {\small $39264$} & {\small $28.00$} & {\small $2.00$} & {\small $26.37$} & {\small $0.53$} & {\small $0.89$} \\
\hspace{1em}{\small Enron email network} (1999-2003) & {\small $87273$} & {\small $1148072$}   & {\small $85.14$} & {\small $14.54$}& {\small $71.05$} & {\small $0.30$} & {\small $0.99(*)$}\\
\hspace{1em}{\small e-mails network European institution} & {\small $265214$} & {\small $420045$}   & {\small $76.02$} & {\small $6.09$}& {\small $70.30$} & {\small $0.30$} & {\small $0.84(*)$} \\
\hline
\textbf{Citation} &   &   &  & &  & \\
\hspace{1em}{\small Citations to Milgram's 1967 paper (2002)} & {\small $395$} & {\small $1988$}  & {\small $10.48$} & {\small $10.48$} & {\small $4.49$} & {\small $1.00$} & {\small $1.00$} \\
\hspace{1em}{\small Articles from Scientometrics (1978-2000)} & {\small $3084$} & {\small $10416$}   & {\small $10.32$} & {\small $8.32$}& {\small $5.28$} & {\small $0.98$} & {\small $1.00$} \\
\hspace{1em}{\small Citation network DBLP} & {\small $12591$} & {\small $49743$}   & {\small $21.50$} & {\small $16.82$} & {\small $8.45$} & {\small $0.87$} & {\small $1.00$} \\
\hline
\textbf{Social} &   &   &   & &  &  \\
\hspace{1em} {\small Hyper-network of 2004 US election blogs} & {\small $1224$} & {\small $19025$}   & {\small $45.37$} & {\small $10.95$} & {\small $34.95$} & {\small $0.72$} & {\small $0.98$}\\
\hspace{1em} {\small Reply network of the news website Digg} & {\small $30398$} & {\small $87627$} & {\small $15.92$} & {\small $6.56$} & {\small $10.18$} & {\small $0.61$} & {\small $0.97$}\\
\hspace{1em} {\small Trust network from the website Epinions} & {\small $75879$} & {\small $508837$}   & {\small $123.00$} & {\small $16.47$} & {\small $106.96$} & {\small $0.13$} & {\small $0.80(*)$} \\
  \hline  
\end{tabular}
\caption{We report values of metrics associated to non-normality for a selection of well-studied real-world networks, whose numbers of nodes and links span several orders of magnitude. A more complete table is available in the SM. All networks are weighted and directed, and appear to have a significant non-normality.  Their adjacency matrix $\mathbf{A}$ satisfies $\mathbf{A}\mathbf{A}^T\neq \mathbf{A}^T\mathbf{A}$ and they posses a positive numerical abscissa ($\omega$) - i.e. they are reactive - which is much larger than the corresponding spectral abscissa ($\alpha$){, indeed almost all the values in the column $\omega-\alpha$ are strongly positive}. Moreover the $\epsilon$-pseudospectral abscissa, $\alpha_{\epsilon}$ {(see Table~\ref{box:NN})}, is positive for the value $\epsilon=10^{-1/2}$. The normalised Henrici's departure from normality, $\hat{d}_F$ {(see Table~\ref{box:NN} and Methods)}, is also often very large. We observe that, in the case  of very large networks, denoted with a $(*)$, we have only been able to  provide an upper bound for this index, by computing the $1000$ largest eigenvalues. We also report the structural measure of {asymmetry} $\Delta$ which is correlated with $\hat{d}_F$ (see  {text and} Fig.\ref{fig1b} panel (c)). Additional details are provided in Figs. 4 and~5 in the SM.} 
\label{tab:table}
\end{table*}

Although a system can initially be close to an asymptotically stable equilibrium, it can leave this state even when a moderate external perturbation occurs due to non-normality~\cite{spectra}. This effect is even more striking once one includes stochastic forces to the model, e.g. exogenous or demographic perturbations due to the surrounding environment~\cite{gardiner}, as they may push even a linear, stable system out of equilibrium when it is non-normal. Again, this behaviour can not be properly captured {neither} by the spectrum of the linear model, nor can it be  described by the numerical abscissa, only determining the short time behaviour of the dynamics. To describe the long time consequences of such perturbations, another important tool is the {\em pseudospectrum} {$\sigma_\epsilon(\textbf{M})=\{\sigma(\textbf{M}+\textbf{E}):||\textbf{E}||\leq\epsilon\}$, for $\epsilon>0$}~\cite{spectra} {, from which one can compute the $\epsilon$-{\em pseudospectral abscissa} replacing somehow the role of the spectral abscissa and eventually the {\em Kreiss constant} $\mathcal{K}$ which provides a direct measure of the size of the transient amplification (see Table~\ref{box:NN})}. By definition, a complex number $z$ is an eigenvalue of $\textbf{M}$ if a bounded inverse of $z \textbf{I} -\textbf{M}$ does not exist. The pseudospectrum is based on a less strict  definition and defines regions of the complex plane where {$||(z \textbf{I} -\textbf{M})^{-1}||$} is larger than a prescribed {positive} number $\epsilon^{-1}$.
{By its very first definition}, the  pseudospectrum defines regions of the complex plane where eigenvalues of a matrix can be found due to a small perturbation, $\textbf{M} + \Delta \textbf{M}$, with $||\Delta \textbf{M}|| < \epsilon$. Such perturbations lead to small variations of the spectrum in the case of normal matrices, but they can become much more important in the case of non-normal matrices. In particular, even small perturbations can make a linearly stable system unstable. 
Note that this effect may have important practical consequences for networks, as the precise value of edge weights is often unknown~\cite{liu2011controllability} and empirical measurements of networks are prone to missing edges~\cite{liben2007link}.  

As we have discussed, non-normality may strongly affect linear and nonlinear dynamical systems on networks and, more generally, their behaviour. The  contributions of this work are manifold. First, we show that a strong non-normality is  widespread in  complex networks empirically observed in {a} variety of domains. As a second step, we reveal the organisation behind non-normality and show that non-normality is associated with a combination of absence of cycles \cite{johnson2017looplessness}, low reciprocity \cite{squartini2013reciprocity} and  hierarchical organisation \cite{Hvsmotifs}. We also propose a simple model for growing networks based on preferential attachment reproducing our observations. Finally, we consider in detail  a Lotka-Volterra model applied to a real-world network and show that the use of network metrics for non-normality helps to understand the dynamics of the system.

\begin{figure*}[t]
\centering
\includegraphics[scale=0.4]{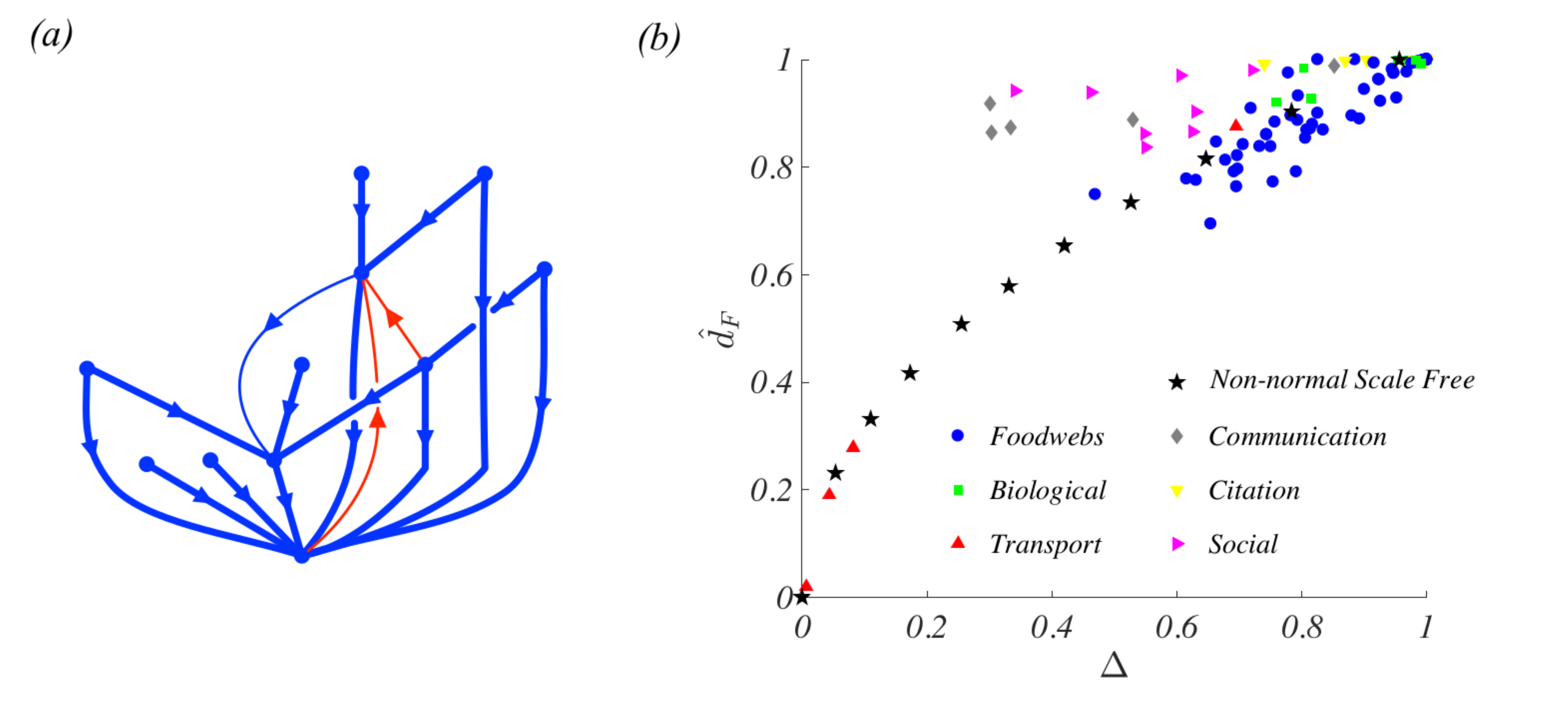}
\caption{\textbf{Structure of non-normal networks.} $\textbf{(a)}$ DAG (blue links) embedded in a weighted non-normal network, with red links corresponding to entries in the lower triangular part of the adjacency matrix (once nodes are reordered). The thickness of a link is proportional to its weight. {$\textbf{(b)}$} Normalised Henrici's departure from normality versus the structural measure of asymmetry $\Delta$, for the networks of Table I SM and for the nSF model.}
\label{fig1b}
\end{figure*}

\vspace*{1.cm}
%\section*{Results}
%\subsection*{Non-normal networks: empirical data and generation models}
%\label{sec:2}
\noindent{\Large \textbf{Results}}\vspace*{.2cm}\\
\noindent{\textbf{Non-normal networks: empirical data and the shape of non-normality.}}
As a first step, we have considered a large set of directed, real-world networks from different disciplines, including biology, sociology,  communication, transport and many more. Results reported in Table \ref{tab:table} (see also the more complete Table {presented} in SM) show values of standard measures of non-normality, including the numerical abscissa, the $\epsilon$-pseudospectral abscissa and the normalised Henrici's departure from normality, {$\hat{d}_F(\mathbf{M})=d_F(\mathbf{M})/{||\mathbf{M}||_F}$}, all revealing that the networks present a strong non-normality.

As a next step, we investigate the type of network organisation associated to non-normality.
The directedness and low reciprocity of a network are necessary conditions for non-normality, but they are by no means sufficient. For instance, a $k-$regular directed ring, whose adjacency matrix is  circulant, is normal due to its rotational symmetry~\cite{nonnormal}. The condition $\textbf{A} \textbf{A}^{T} \neq \textbf{A}^{T} \textbf{A}$  is instead satisfied when the network is hierarchical, that is when nodes have a rank and edges with a strong weight tend to flow from nodes with a small rank to nodes with a high rank (or vice versa). 
Such {organisations} are known to be prevalent in different types of networks \cite{hierarchy,RWhierarchy,Hvsmotifs,de2018physical}, for instance through the concepts of dominance hierarchies in social ecology, trophic levels in foodwebs  and social status in social networks. The inequality becomes maximum when the network is a directed acyclic graph (DAG), such that the matrix takes an upper triangular form after proper relabelling of the nodes. Again, DAGs find several applications, for instance in the case of citation or causal networks. 
Based on these intuitions, we estimate the level of hierarchy of a real-world network as follows. Given an adjacency matrix, we search for the best nodes ordering such that the total weight in the upper-triangular part of the matrix is maximal (See Methods), thus allowing to identify the DAG of maximum weight embedded in the network. 
A simple measure of {asymmetry} is then the unbalance $\Delta$ between the number of entries in the upper and lower triangular part of the relabelled adjacency matrix $\tilde{\textbf{A}}$, that is $\Delta:=\left| K^<-K^>\right| / K$, where $K^<=\sum_{i<j} \tilde{A}_{ij} $, $K^>= \sum_{i>j} \tilde{A}_{ij}$ and $K=K^<+K^>$.
$\Delta$ thus measures the asymmetry of the adjacency matrix after relabelling and it provides a structural indicator of non-normality that can be compared with standard spectral measures, 
such as the {normalised} Henrici's departure from normality. In panel (c) of Fig.~\ref{fig1b}, we show that the normalised Henrici's index strongly correlates with $\Delta$ across a variety of real-world networks, hence reinforcing the connection between structural hierarchy and dynamical non-normality {(see also Methods)}. 
Finally, note that non-normality and  a hierarchical structure are {global properties} of a network. As an illustration, we {have} consider{ed in the SM} the case of networks built from different combinations of the same {constituting} blocks, or 
 motifs ~\cite{alon,Milo2004,Hvsmotifs,Gorochowski} {and we observe that different levels of non-normality can emerge from the combination of two directed motifs, from strong non-normality and a DAG structure for the whole graph, to weakly non-normal patterns where the presence of {a cycle} prevents the dynamical flow to accumulate on a small number of nodes}.

\begin{figure}[t]
\centering
\includegraphics[width=\textwidth]{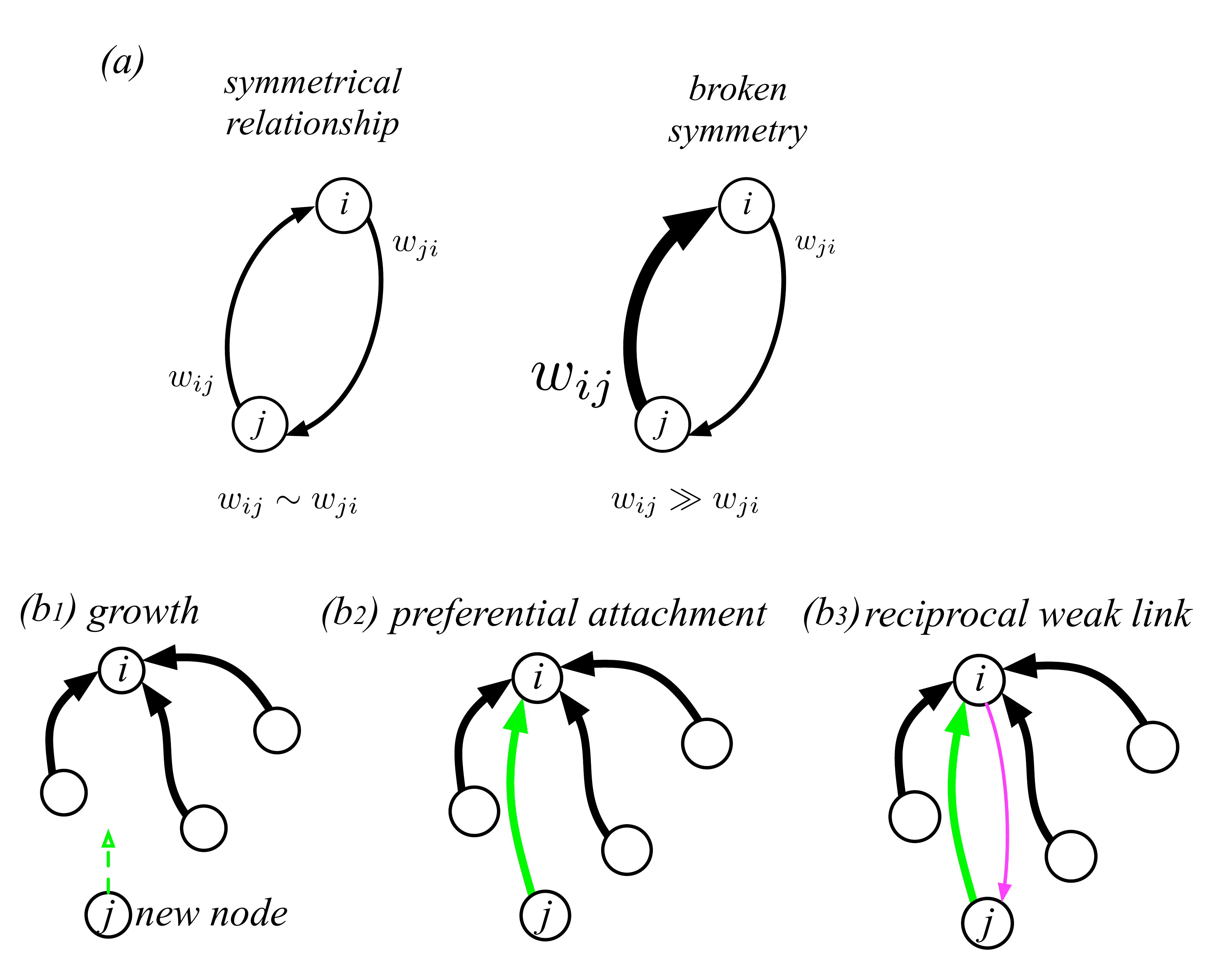}
\vspace{-.5cm}
\caption{\textbf{Non-normal Scale-Free network.} 
\textbf{(a)} {\em local breaking of reciprocity}: if $j$'s influence over $i$ is larger than the reciprocal one,  the pairwise relation is non-reciprocal and presents a broken symmetry. \textbf{(b)} {\em generating model of nSF network}: once a new node enters into the system (dashed green arrow in $b_1$), it establishes, with higher probability, a link pointing toward the node with larger {in-}degree, $i$, (solid green link in $b_2$). With a lower probability, the latter, i.e. the hub, can create a weaker link pointing in the reciprocal direction (thin magenta link in $b_3$).}
\label{fig1}
\end{figure}   

\begin{figure*}[t]
\centering
\includegraphics[scale=0.35]{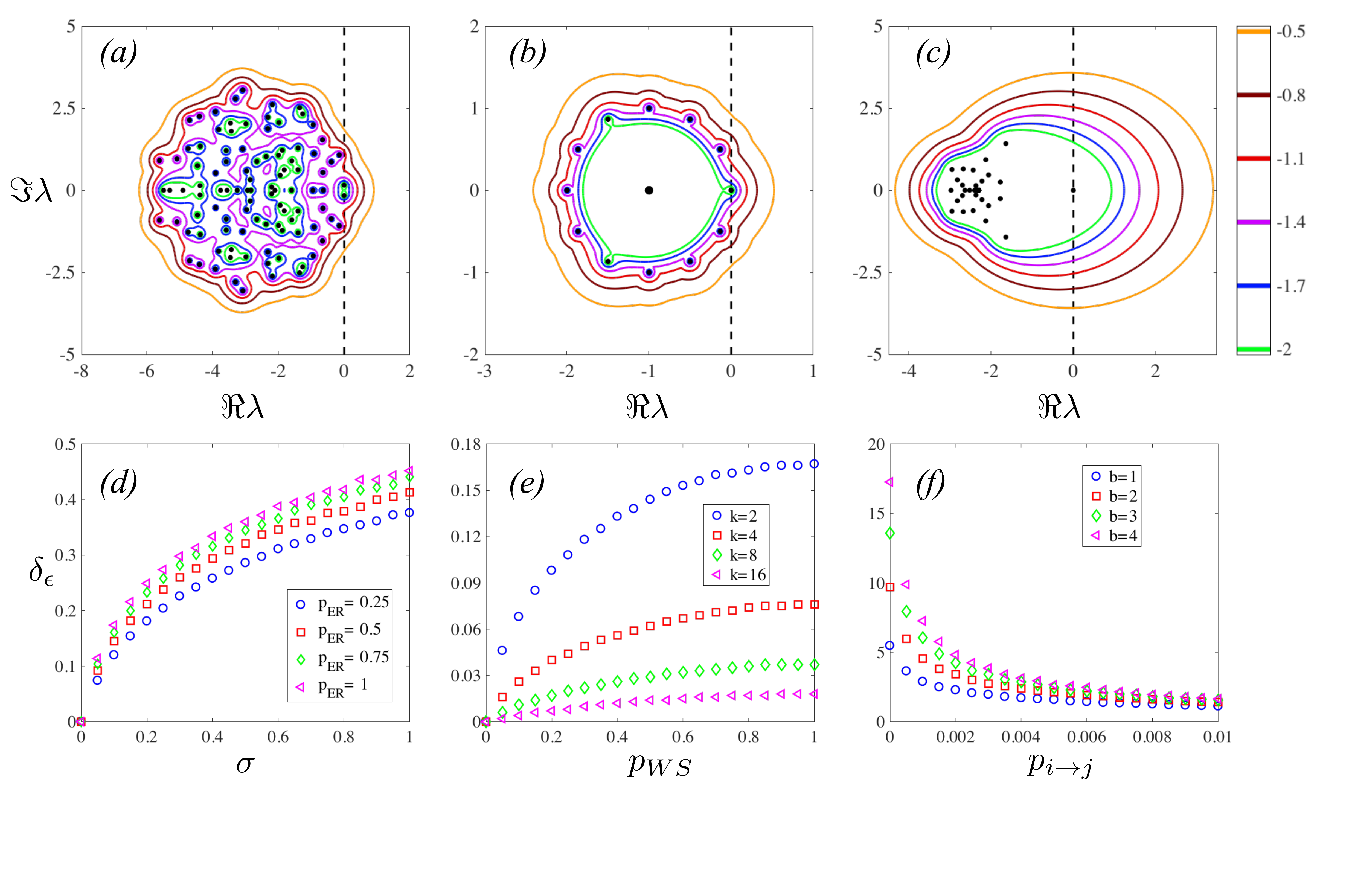}
\vspace{-1.5cm}
\caption{\textbf{ Models of non-normal networks}.  Spectra and pseudospectra {(computed using the software \textit{Eigtool}~\cite{eigtool})} of {non-normal} models (upper panels): $\textbf{(a)}$ Erd\H{o}s-R\'enyi (ER) with links probability $p_{ER}=0.1$ and weights from a normal distribution $\mathcal{N}(0,1)$, $\textbf{(b)}$ (unweighted) Watts-Strogatz (WS) with initial number of neighbour nodes $k=2$ and rewiring probability $p_{WS}=1$, $\textbf{(c)}$ non-normal scale free network (nSF) with probability of backward links {$p_{i\rightarrow j}=0.001$ for $j>i$} and weights from a uniform distribution, $\mathcal{U}[0,1]$. The colour bar, on the right, represents the different levels of $||\textbf{E}||$ {in log scale e.g. $\epsilon=10^x$ where $x$ is the numerical value reported on the bar}. {Visual inspection reveals that the pseudospectrum of diverse network models are affected differently by $\epsilon$ and, in particular, that its size is increased more significantly for the nSF model. In the lower panels, we quantify how the size increase of the pseudospectrum affects the stability of its network as follows. We consider the difference between the $\epsilon$-pseudoabscissa of $\mathbf{A}$ and that of its Hermitian part, $\delta_{\epsilon}=\alpha_{\epsilon}(\mathbf{A})-\alpha_{\epsilon}(H(\mathbf{A}))$, hence measuring how the non-Hermitian aspect of the system affects its dominant eigenvalue. This quantity is always positive}, as the pseudospectrum  of a non-normal matrix is larger than that of a normal one~\cite{spectra}, for any given fixed $\epsilon>0$. $\textbf{(d)}$ ER with weights from a normal distribution $\mathcal{N}(0,\sigma)$ {for several values of the variance} and different links probabilities, $p_{ER}$. $\textbf{(e)}$ (unweighted) WS for different rewiring probabilities $p_{WS}$ and different initial number of neighbour nodes $k$. $\textbf{(f)}$ nSF with varying backward links probability $p_{i\rightarrow j}$ and different upper bound, $b$, of the uniform distribution from which the weights are chosen, $\mathcal{U}[0,b]$. In each case, the networks are composed of $100$ nodes and the adjacency matrices have been diagonally shifted with their respective spectral abscissa $\alpha(\mathbf{A})$, in order to set the real part of the maximum of each spectrum exactly at $0$. {For the lower panels the value of $\epsilon$ has been set to $10^{-0.5}$.}}
\label{fig2}
\end{figure*}

\noindent{\textbf{Mechanistic model.}}
Here we propose a simple mechanistic model, denoted by non-normal Scale-Free network (nSF), leading to the formation of  networks with tuneable levels of non-normality. The model builds on the seminal ideas of Price \cite{price}, later leading to the family of preferential attachment models \cite{scale-free}. As discussed before, critical ingredients of the model should be the low reciprocity of the directed edges and the presence of a hierarchical structure. We thus consider a growing network where, at each step, a new node $j$ draws a directed edge to a previously existing node $i$, with a probability proportional to its in-degree. Note that this type of process is expected to lead to the formation of power-law in-degree distributions, but this is not our concern here. Node $i$ also has the possibility to reciprocate and to create a link directed at $j$, as in~\cite{zlatic2011model}. 
Asymmetry can be included in different ways, either by endowing edges with a weight and imposing that $0\leq w_{ji} \ll w_{ij}$, or by considering unweighted edges and assuming that the reciprocal edge is created with a probability $p_{i\rightarrow j}{\ll}1$. Hierarchy is then induced by the  ordering of the nodes in terms of their arrival time. As expected, the stronger the inequalities, the stronger the non-normality of the resulting networks. We have investigated the non-normality of the resulting networks and found a striking similarity with the relation between $\Delta$ and the Henrici's departure from normality observed in real-world networks, as shown in Fig.~\ref{fig1b}. Note that we have also considered variants of other classical network models, such as Erd\H{o}s-R\'enyi (ER)~\cite{ER} and Watts-Strogatz (WS)~\cite{watts} (see SM) but their lack of hierarchical structure prevents the formation of strong non-normality, as can be seen in their pseudospectral properties (see Fig.~\ref{fig2} and SM).

\begin{figure*}[t]
\centering
\includegraphics[scale=0.36]{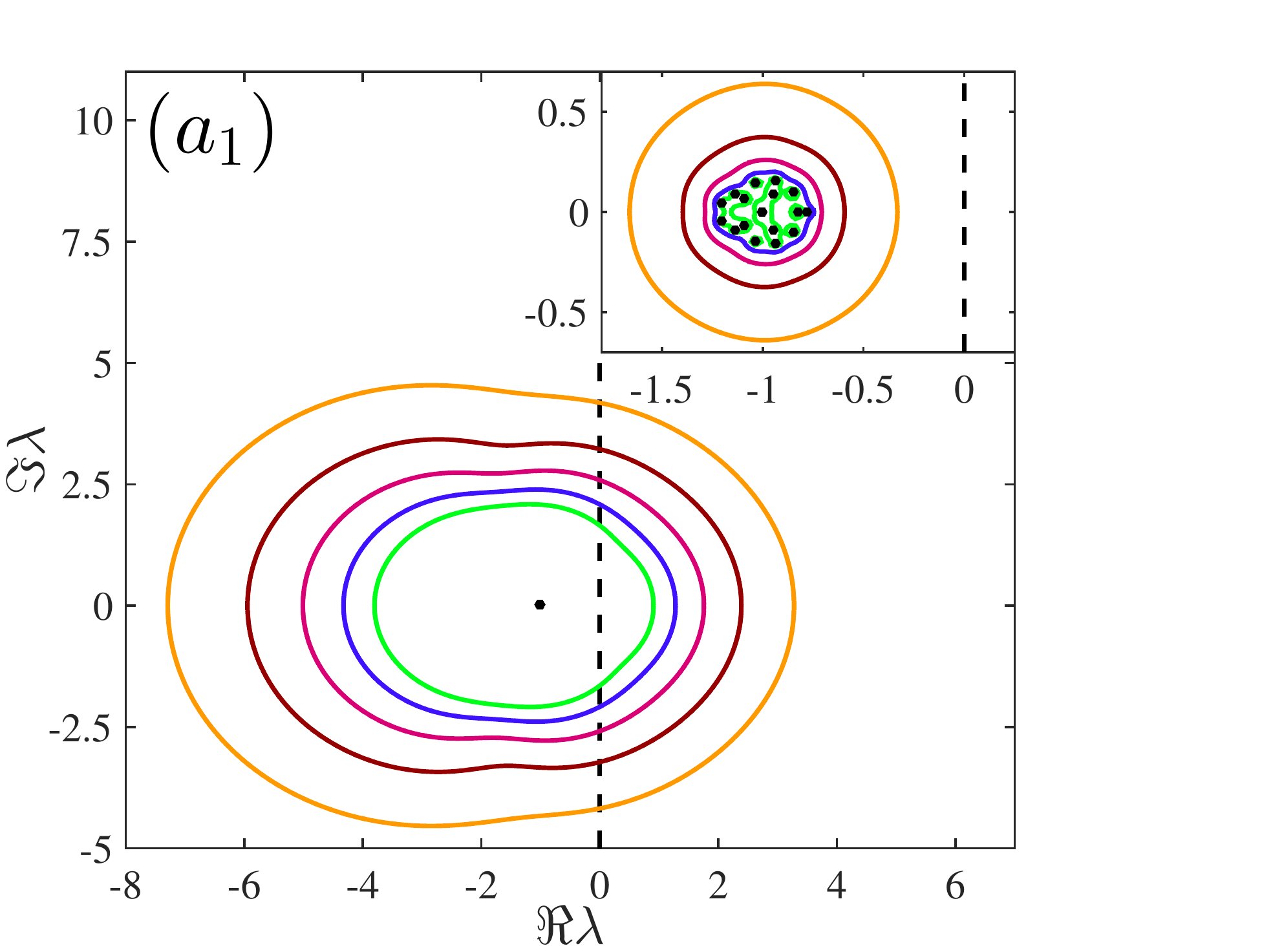}\hspace*{-1.35cm}\includegraphics[scale=0.36]{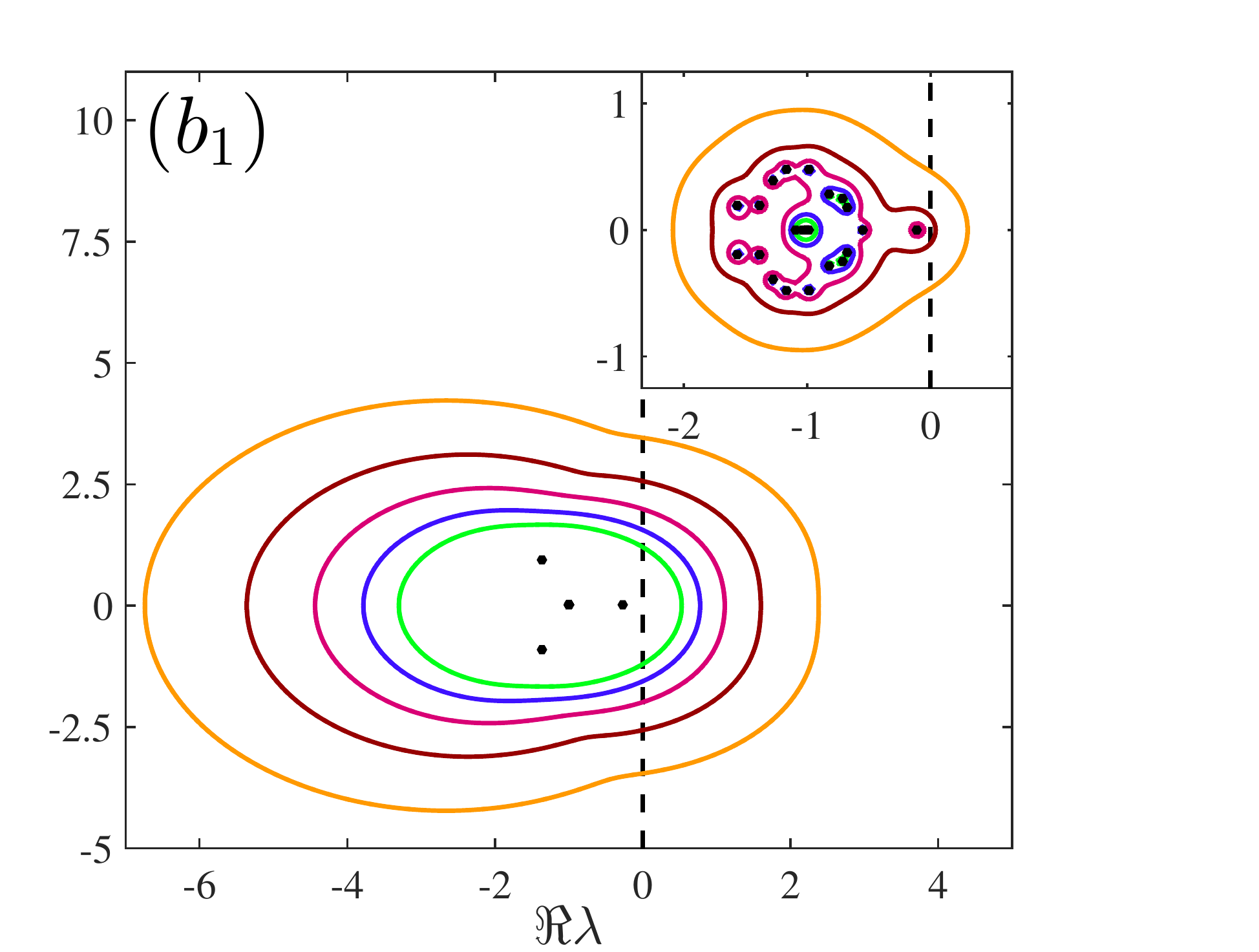}\hspace*{-1.35cm}\includegraphics[scale=0.36]{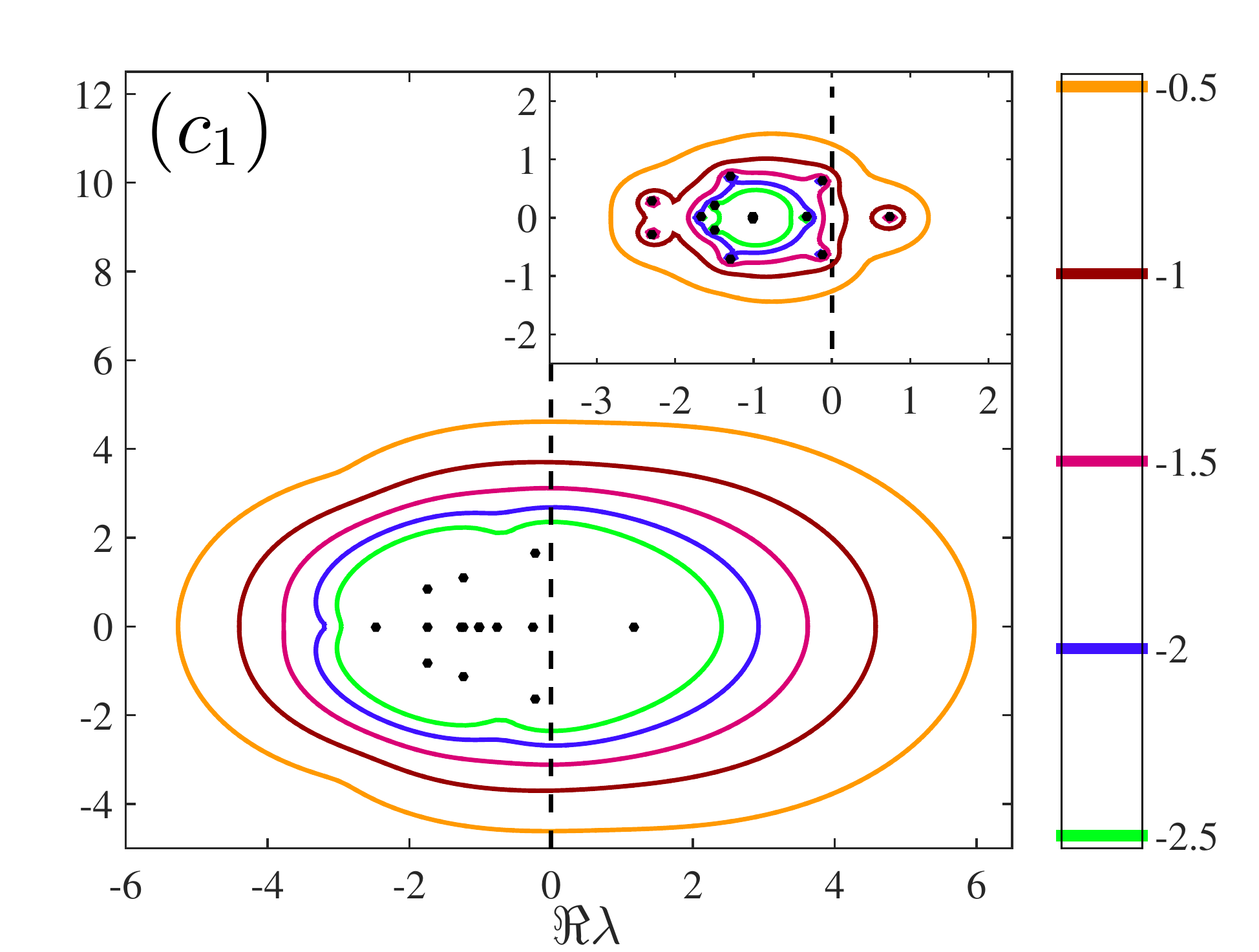}\\
\includegraphics[scale=0.33]{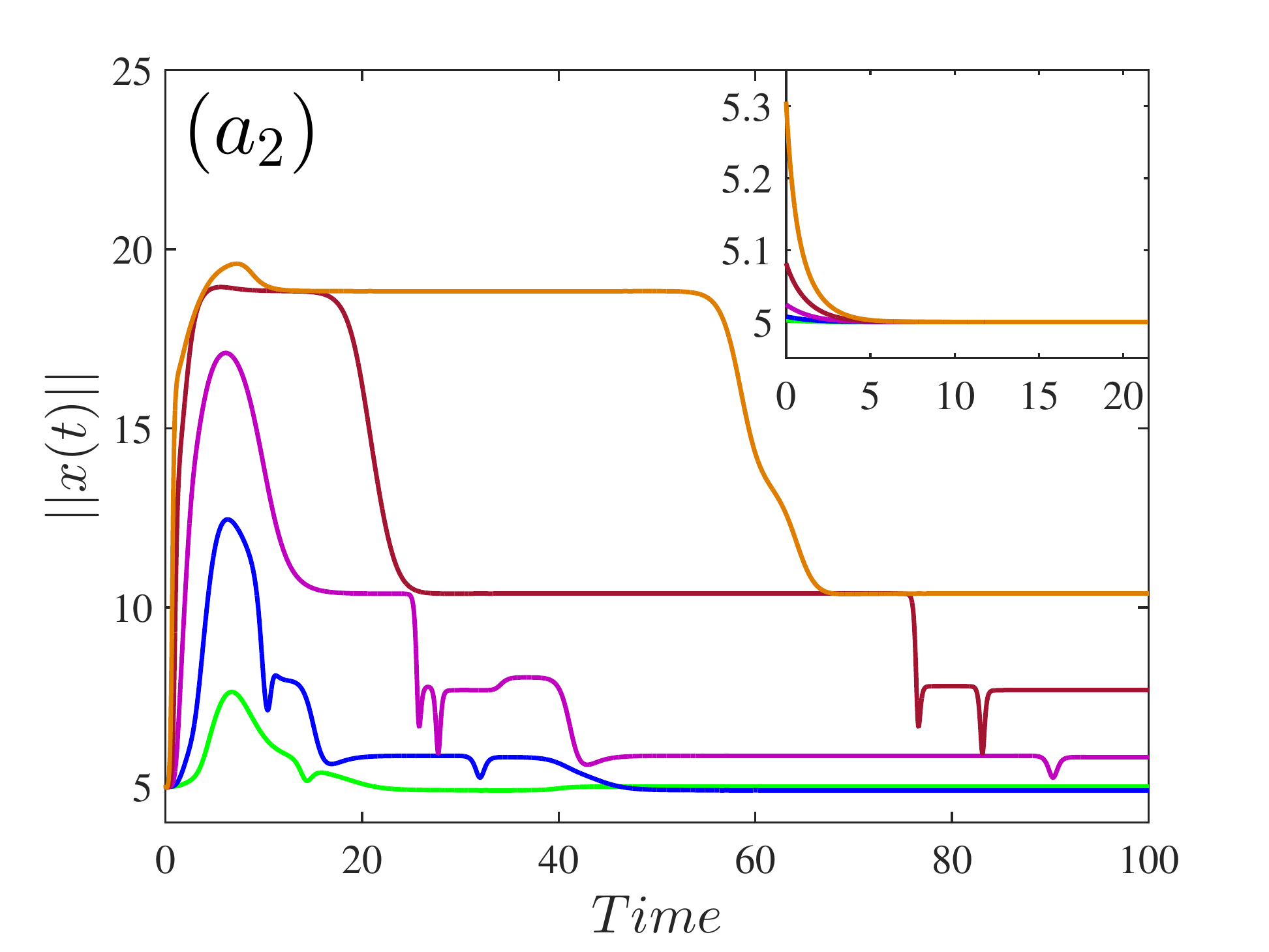}\hspace*{-0.47cm}\includegraphics[scale=0.33]{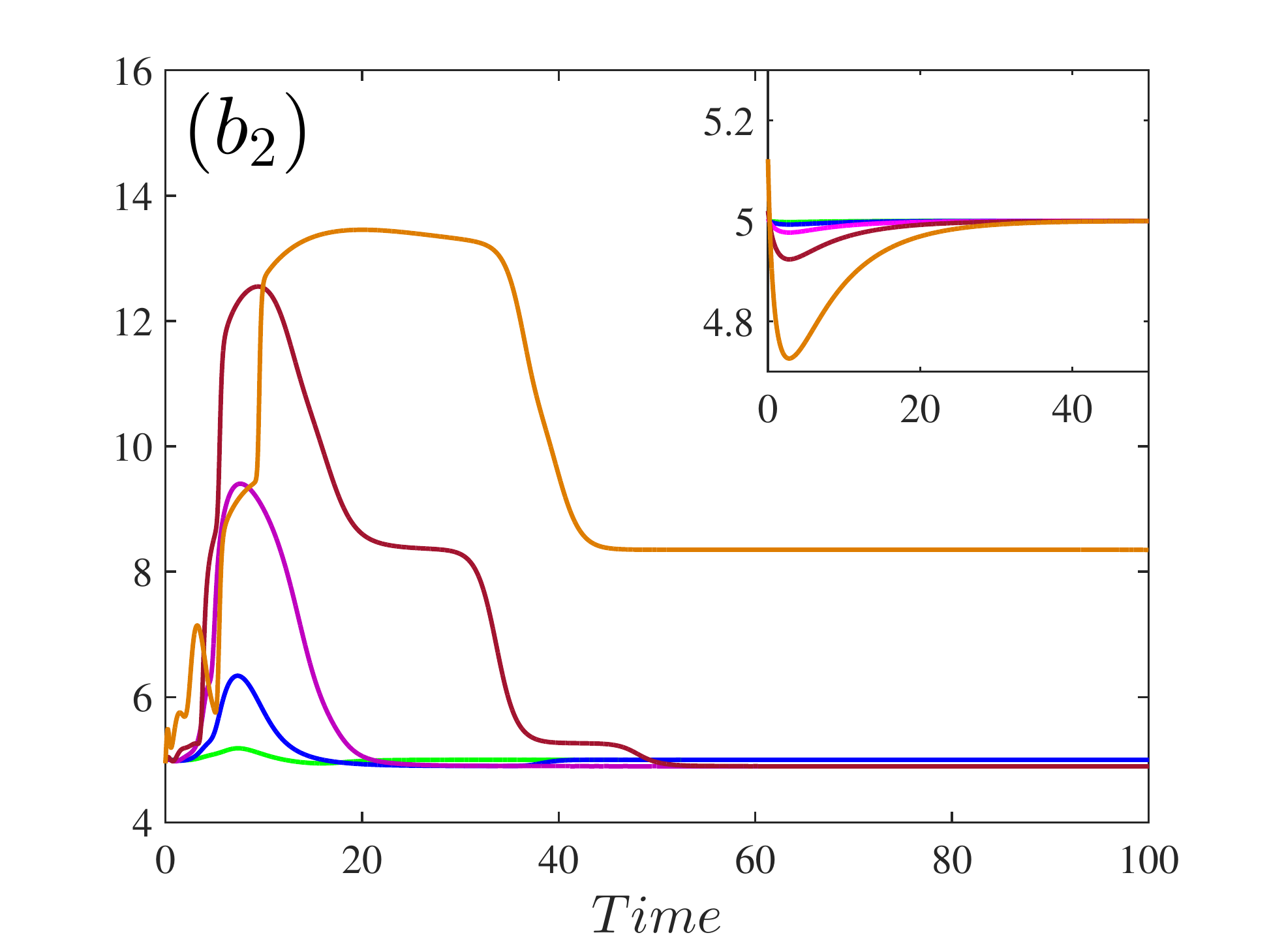}\hspace*{-0.47cm}\includegraphics[scale=0.33]{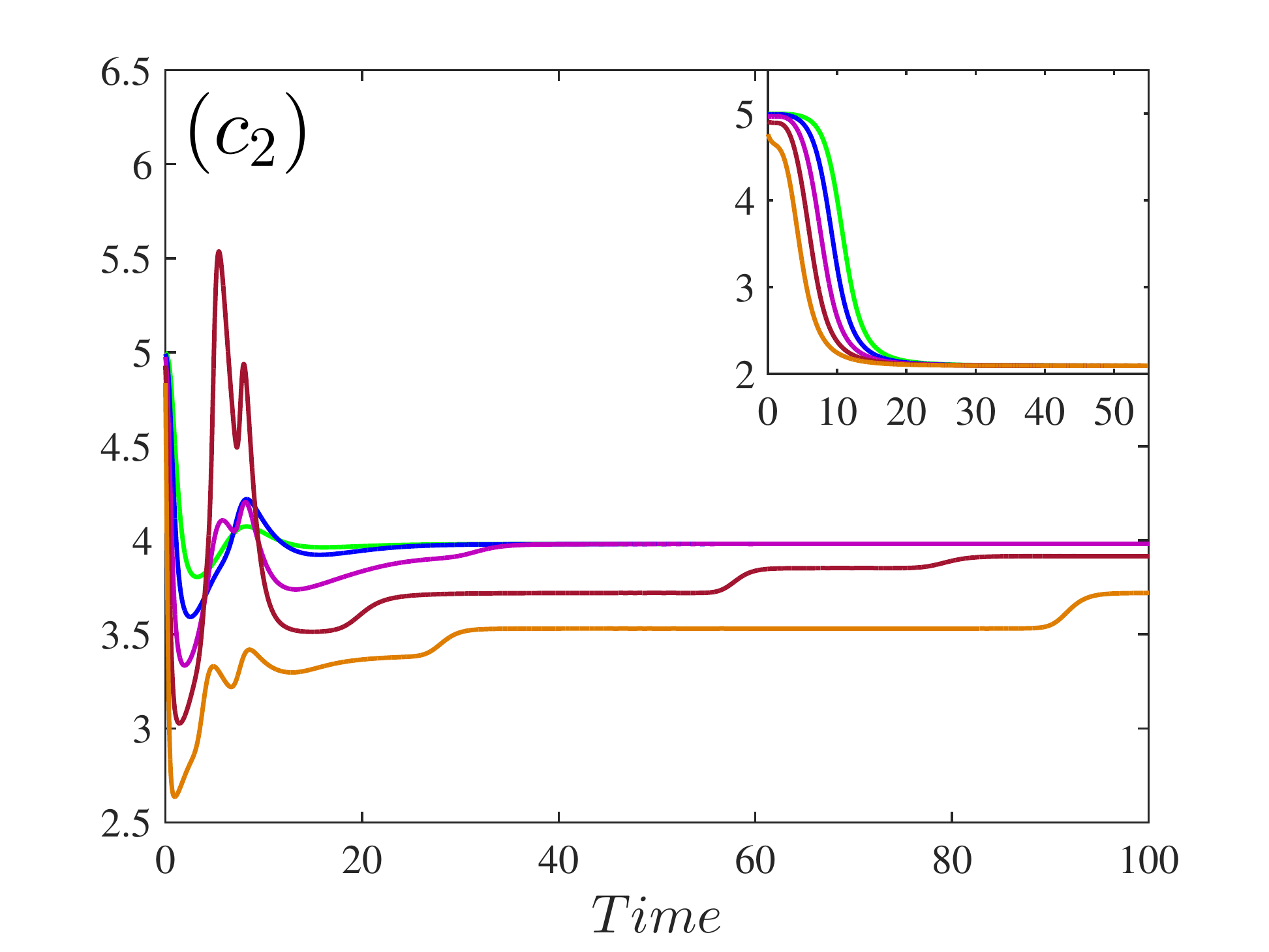}
%\vspace{-2.cm}
\caption{\textbf{Generalised Lotka-Volterra model:} $\dot{x}_i=x_i(r_i-s_ix_i+\sum_{j\neq i} M_{ij}x_j)$\textbf{.} We consider an ecosystem composed of {$25$} species. For the sake of simplicity, the intra-species interactions are all set equal, $s_i=1\; \forall i$, and $\mathbf{M}$ is the {(weighted and signed)} adjacency matrix of a nSF  network for the structured case (main panels) or a random matrix (insets) whose weights are drawn from a normal distribution $\mathcal{N}(0,1/5)$. In the structured cases the strengths in the upper triangular part of the matrix $\mathbf{M}$ are {$15$} times larger than in the lower one,  thus enhancing  non-normality as  can be seen from the pseudospectrum levels {(computed using the software \textit{Eigtool}~\cite{eigtool})}. In the upper panels, we show the Master Stability Function close to the asymptotically stable equilibrium point, {based on the use of} the pseudospectra. The corresponding dynamical evolution is shown in the lower panels {where different colours correspond to different levels of $\epsilon$ in log scale (as in the upper panels), initial conditions have been uniformly randomly chosen}. Different cases are considered depending on the signs of the interaction strengths: $\textbf{(a)}$ competition  $(-/-)$, $\textbf{(b)}$ prey-predator $(-/+)$ and $\textbf{(c)}$ mutualism $(+/+)$. We  observe that, even if the system is asymptotically stable (panels ($a_1$) and ($b_1$)), the $\epsilon$-pseudospectral abscissa is positive for sufficiently large  $\epsilon$,  thus inducing an unstable system behaviour if the perturbation (in the adjacency matrix and/or in the initial conditions) is strong enough {(panels ($a_2$) and ($b_2$))}. Yet, the unstructured systems still converge to the homogeneous equilibrium (see insets in panels ($a_2$) and ($b_2$)). Overall, this effect is more pronounced in the structured systems than in the random ones, as the $\epsilon$-levels are much larger in the former case, for a fixed value of $\epsilon$.}
\label{fig3}
\end{figure*}

\vspace*{.5cm}

%\subsection*{Application to the stability of complex ecosystems}
%\label{sec:4}
\noindent{\textbf{Application to the stability of complex ecosystems.}}
%~\footnote{Tao \& Vu~\cite{tao} proved that the eigenvalues of a {$n\times n$ matrix, whose entries are i.i.d. random variables drawn from a} distribution with $0$ mean and variance $1$, lie in a disc of radius $\sqrt{n}$ with probability $1$ for large enough $n$.}}. %\cite{footnote1}. %\footnote{Actually, in the original model of May \cite{may} the diagonal elements were equal to some negative value $-a$, $a>0$ representing the death rate of the species. This is the reason why the system is generally stable for $a<\sqrt{n}$ and unstable otherwise.}.
% Consequently, an increase of the numbers of the interconnected components will transit the system to instability. It has been recently shown that such models become more vulnerable, and thus the systems less stable, when the competitiveness among species decreases \cite{zoology}.
The hierarchical structure of non-normal networks allows to introduce  interesting connections with dynamical systems. 
Here, we focus on stability, a central concept to understand  the emergence of collective phenomena \cite{master}. The importance of network structure for stability is well-established since the seminal works of May \cite{may,allesina} in the context of ecology. For instance, choosing the interaction strengths from a normal distribution {$\mathcal{N}(0,\sigma)$}, May proved that an ecosystem  loses its stability above a critical size, as consequence of the circular law~\cite{tao}. 
To  understand the interplay of non-normality and dynamics, we analyse the Master Stability Function~\cite{master,newman} {which is a general tool allowing to infer about the (in)stability of a networked dynamical system; it often relies on the use of the spectrum of some suitable linearisation, while hereby condition for (in)stability are determined through to the pseudospectrum of the linearised system~(\ref{eq:GLV}). The latter representing the} Generalised Lotka-Volterra (GLV) model, popular to understand competition and mutualism among interacting species \cite{may,allesina,zoology}. {The set of equations governing the dynamics of trophic interactions is given by~\cite{zoology}:
\begin{equation}
\frac{dx_i}{dt}=x_i\left(r_i-s_ix_i+\sum_{j\neq i} {M}_{ij}x_j\right), \; \forall i=1,2,\dots N.
\label{eq:GLV}
\end{equation} 
Here $r_i$ are the intrinsic rates of i) birth if $r_i>0$, meaning that species $i$ can reproduce itself in absence of other species and in abundance of resources; ii) death  if $r_i<0$ in the sense that the population of species $i$ will decline in absence of other species (e.g, preys). The positive constants $s_i$  represent the finite carrying capacity of the ecosystem (limited resources) and prevent the species $i$ to grow indefinitely. An important role is played by the community matrix $\mathbf{M}$ whose entries ${M}_{ij}$ (resp. ${M}_{ji}$) represent the influence of species $j$ on $i$ (resp.  $i$ on $j$). We also assume that ${M}_{ii}=0\, ,\forall\;i$, namely the community matrix describes only  inter-species interactions, and intra-species interactions have been cast into $s_i$. In the following, we  adopt the method of Chen and Cohen~\cite{chen}, as already done in the literature~\cite{allesina,zoology}. More precisely, we hypothesise the existence of a positive equilibrium solution $\mathbf{x}^*$, that without loss of generality can be assumed to be of the form $x_i^*=1$, for all $i$, after a suitable choice of the growth/death rates $r_i$.
At this point the Master Stability Function of the GLV model depends solely on the spectrum (pseudospectrum) of the matrix $\mathbf{M}-\textit{diag}(s)$, namely the community matrix from which we remove the matrix whose diagonal contains the inter-species strengths $s_i$. The problem is hence mapped to a framework where  stability directly depends on species interactions.

The vulnerability of the system is clearly visible in Fig.~\ref{fig3} where the spectra {(black dots)} shift from the left to the right of the imaginary axis once mutualism increases. Although the system remains clearly stable for a strong competitive setting (panels (a) and (b)), we can observe {(coloured curves)} that the $\epsilon$-{\em pseudospectral abscissa} (see~\cite{spectra} and {Table~\ref{box:NN}}) is positive and larger for structured systems than for random ones~\cite{allesina}. {To represent the former systems we used the nSF networks obtained with the generation model previously presented which result to be very similar to the real trophic relations}. This implies that the system can easily be destabilised by (relatively) small fluctuations due to demographic, thermal or endogenous noise, always present in the surrounding environment and that are amplified due to the non-normality (see SM). This remark can have important consequences in the understanding of the problem of coexistence of multiple species in a harsh competitive environment, {e.g. in the case of the paradox of the plankton~\cite{plankton} for which field observations are at odds with the Principle of Competitive Exclusion \cite{allesina}}.

\vspace*{1cm}
%\section*{Discussion and conclusions}
\noindent{\large \textbf{Discussion and conclusions}}\vspace*{.2cm}\\
We have shown that a large number of real-life networks are strongly non-normal, and that a characterisation of their properties solely by spectral methods may be misleading. Non-normality induces a strong dependency on fluctuations and needs to be considered with care when performing a linear stability analysis of non-linear systems. Despite that the non-normality is well-studied and that its importance has been recognised in {a} variety of domains, a systematic analysis of its importance and effect in large-scale networks was still lacking. Our first contribution is thus the identification of what appears to be an ubiquitous property of directed networks, but also the introduction of new methods in the toolbox of network science, to generate non-normal networks and capture the effect of non-normality on their dynamics. Potential applications have recently been explored, for instance in pattern formation on networks 
 \cite{nonnormal}, and in epidemic spreading in metapopulation models \cite{nonnormal}.
Overall, these findings emphasise that  non-normality is a critical component of complex systems and that specific tools are necessary to complement standard methods based on eigenspectra, which are prevalent in network science. More specifically, this new {perspective} may shed light to explain the diversity of species in ecosystems~\cite{diversity}, the origin of cascade failures in power grids~\cite{havlin} or the spread of epidemics in mobility networks~\cite{colizza}, just to mention few possible applications.  

\vspace*{1cm}
\noindent{\large \textbf{Methods}}\vspace*{.2cm}\\
\noindent{\textbf{{Measures of non-normality.}}} A real matrix $\textbf{M}$ is said to be \textit{non-normal} if it is not diagonalisable by a unitary matrix, namely its eigenvectors are not orthogonal to each other~\cite{spectra}. The \textit{numerical abscissa} has been introduced in population dynamics \cite{ecoresilience} with the name of \textit{reactivity} and it is defined by $\omega(\textbf{M})=\sup\sigma\left(H(\textbf{M})\right)$, being $H(\textbf{M})=(\textbf{M}+\textbf{M}^T)/2$ the Hermitian part of $\textbf{M}$. This is indeed a very natural concept, however it doesn't allow to compute the maximum amplification of the initial conditions exhibited by linear stable non-normal systems; for this reason one has to recur to the {\em pseudospectrum}~\cite{spectra}, $\sigma_{\epsilon}(\mathbf{M})$, being defined for all $\epsilon >0$ as the spectrum of the perturbed matrix $\textbf{M}+\textbf{E}$, for any matrix $||\textbf{E}||\leq\epsilon$. From the $\epsilon$-pseudospectral abscissa, $\alpha_{\epsilon}(\mathbf{M})=\sup\Re\sigma_{\epsilon}(\mathbf{M})$, we can obtain the Kreiss constant, $\mathcal{K}(\textbf{M})=\sup_{\epsilon>0} \alpha_{\epsilon}(\textbf{M})/\epsilon$, and eventually the lower bound on the orbit size
\begin{equation}
\sup_{t\geq 0} ||{\mathbf{x}(t)}||\geq \mathcal{K}(\textbf{M})  ||{\mathbf{x}(0)}||\, .
\end{equation}
Let us observe that the latter provides a straightforward bound on the amplification envelope defined in~\cite{ecoresilience}. Moreover $\mathcal{K}(\textbf{M})$ is more informative that reactivity, in fact a stable system can exhibit a small amplification even if $\omega(\textbf{M})>0$ is very large.

The Henrici's index is based on the observation that the Frobenius norm of a normal matrix is given by $||\mathbf{M}||_F^2=\mathit{tr}(\mathbf{M}^T\mathbf{M})=\sum_{i=1}^n|\lambda_i|^2$ where $\lambda_i$ are the eigenvalues of the matrix; one can thus define the {\em Henrici's departure from normality}~\cite{spectra} for a non-normal matrix $\mathbf{M}$ by:
\begin{equation}
d_F(\mathbf{M})=\sqrt{||\mathbf{M}||_F^2-\sum_{i=1}^n|\lambda_i|^2}\, .
\end{equation}
It attains its minimum once the matrix is indeed normal and then increases as long as the matrix deviates from normality. To compare systems with different sizes we define the normalised index $\hat{d}_F(\mathbf{M})=d_F(\mathbf{M})/{||\mathbf{M}||_F}$.

For a generic matrix with binary (resp. positive entries), one can define the unbalance between the number (resp. the total sum) of entries in the upper and lower triangular part, or using the language of networks 
\begin{equation}
\label{eq:Delta}
\Delta(\mathbf{M}):=\frac{\left|\left(\sum_{i<j} \tilde{M}_{ij} - \sum_{j<i} \tilde{M}_{ij}\right)\right|}{\left(\sum_{i<j} \tilde{M}_{ij} + \sum_{j<i} \tilde{M}_{ij}\right)}\, ,
\end{equation}
where $\tilde{M}_{ij}$ are the entries of the final relabelled matrix.

While it can be relatively easy to determine a DAG, and compute $\Delta$, once we have a drawing of a (small enough) network, this task becomes hard starting from the adjacency matrix or a large network. We observe that the simple operation of relabelling the nodes can change the value of $\Delta$ and the latter increases the larger the number of entries in the adjacency matrix are in upper triangular part, namely links $i\rightarrow j$ where $j>i$.  Having in mind these observations we designed an algorithm aiming at maximising $\Delta$ once couples of nodes are relabelled, i.e. rows and columns of the adjacency matrix are swapped. To overcome the combinatorial difficulty of the problem, we resorted to a Simulate{d} Annealing method~\cite{SA} to get an accurate solution in a relatively short time.}
{A pseudo-code is presented in SM and the generic convergence behaviour of the maximisation process is shown in Fig.~{2 SM}.

\vspace*{.2cm}
\textbf{Supplementary materials}\\
{Supplementary material for this article is available at http://XXXX}\\
{text S1. Non-normal matrices and their pseudospectra}\\
{text S2. Global structure of non-normal networks}\\
{text S3. Models for generation of non-normal networks}\\
{text S4. A glimpse of non-normal dynamics of networked systems}\\
{text S5. Pseudospectra of real non-normal networks}\\
{text S6. Extended table of real non-normal networks}\\
{fig. S1. Time evolution of the norm of the solution of the linear ODE $\dot{\textbf{x}}=\mathbf{Mx}$}\\
{fig. S2. Convergence of the maximisation algorithm}\\
{fig. S3. Henrici's departure from normality versus $\Delta$}\\
{fig. S4. Motifs organisation and non-normality}\\
{fig. S5. The spectra and pseudospectra of non-normal networks}\\
{fig. S6. Pseudospectra for real networks I}\\
{fig. S7. Pseudospectra for real networks II}\\
{table S1. Some figures for real webs}\\
%\vspace*{-.25cm}
\bibliography{Non-normal.bib}
\textcolor{black}{\\
\noindent [45] G. H. Golub and C. F. van Loan, {\em Matrix Computations}, (Johns Hopkins University Press 3ed, 1996)\\
\noindent [46] V. I. Arnol'd, {\em Ordinary differential equations} (Springer, 2006)\\
\noindent [47] C. Chicone, {\em Ordinary differential equations with applications} (Springer-Verlag, 1999)\\
\noindent [48] J. Almuniaa, G. Basterretxeaa, J. Aristeguia, and R. E. Ulanowicz, {\em Estuarine, Coastal and Shelf Science}, \textbf{49}, 363, (1999)\\
\noindent [49]  Pajek datasets \\\texttt{http://vlado.fmf.uni-lj.si/pub/networks/} (2018)\\
\noindent [50] R. E. Ulanowicz, {\em Growth and Development: Ecosystems Phenomenology} (Springer-Verlag, NY, 1986)\\
\noindent [51] M. Monaco and R. Ulanowicz, {\em Mar. Ecol. Prog. Ser.}, \textbf{161}, 239 (1997)\\
\noindent [52] J. Hagy, Eutrophication, {\em Hypoxia and trophic transfer efficiency in Chesapeake Bay}, phd dissertation, College park ed. (University of Maryland, US, 2002)\\
\noindent [53] D. Baird and R. Ulanowicz, {\em Ecol. Monogr.}, \textbf{59}, 329 (1989)\\
\noindent [54] D. Baird, J. Luczkovich, and R. Christian, {\em Estuarine, Coastal, and Shelf Science}, \textbf{47}, 329 (1998)\\
\noindent [55] R. E. Ulanowicz, J. J. Heymans, and M. S. Egnotovich, {\em Annual Report to the United States Geological Service Biological Resources Division }, Ref. No.[UMCES] CBL 00-0176, Chesapeake Biological Laboratory, University of Maryland (2000)\\
\noindent [56] KONECT, \texttt{http://konect.cc/} (2017)\\
\noindent [57]  N. D. Martinez, J. J. Magnuson, T. Kratz, and M. Sierszen, {\em Ecological Monographs}, \textbf{61}, 367 (1991)\\
\noindent [58]  H. C. Kaiser M, {\em PLoS Comput Biol}, \textbf{2}, e95 (2006)\\
\noindent [59]  Shen.-O. SS, Milo R, Mangan S., and Alon U., {em Nat Genet.}, \textbf{31}, 64 (2002)\\
\noindent [60] J. Duch and A. Arenas, {\em Phys. Rev. E}, \textbf{72}, 027104 (2005)\\
\noindent [61] R. M. Ewing, et al. and D. Figeys, {\em Molecular Systems Biology}, \textbf{3} (2007)\\
\noindent [62] J.-F. Rual, K. Venkatesan, T. Hao, T. Hirozane-Kishikawa, A. Dricot, N. Li, G. F. Berriz, F. D. Gibbons, M. Dreze, and N. Ayivi-Guedehoussou, {\em Nature}, 1173 (2005)\\
\noindent [63] Federal Aviation Administration, Air traffic control system command center, \texttt{http://www.fly.faa.gov/} (2017)\\
\noindent [64] T. Opsahl, ``Why anchorage is not (that) important: Binary ties and sample selection'', \texttt{http://wp.me/poFcY-Vw} (2011)\\
\noindent [65] G. Storchi, P. Dell'Olmo, and M. Gentili, "Dimacs challenge 9", \texttt{http://www.dis.uniroma1.it/~challenge9} (2006)\\
\noindent [66]  R. W. Eash, K. S. Chon, Y. J. Lee, and D. E. Boyce, {\em Transp. Res. Record}, \textbf{994}, 30 (1983)\\
\noindent [67] D. E. Boyce, K. S. Chon, M. E. Ferris, Y. J. Lee, K.-T. Lin, and R. W. Eash, {\em Chicago Area Transp. Study}, xii + 169 (1985)\\
\noindent [68] T. Opsahl and P. Panzarasa, {\em Soc. Netw.}, \textbf{31}, 155 (2009)\\
\noindent [69] J. Leskovec, D. Huttenlocher, and J. Kleinberg, CHI 2010, April 10-15, 2010, Atlanta, Georgia US (2010)\\
\noindent [70] J. Leskovec, D. Huttenlocher, and J. Kleinberg, WWW 2010, April 26-30, 2010, Raleigh, North Carolina, US (2010)\\
\noindent [71] J. Leskovec and A. Krevl, {\em SNAP Datasets: Stanford large network dataset collection},\texttt{http://snap.stanford.edu/data} (2014)\\
\noindent [72] B. Klimt and Y. Yang, in {\em Proc. European Conf. on Machine Learning} 217-226, (2004)\\
\noindent [73] J. Leskovec, J. Kleinberg, and C. Faloutsos, {\em ACM Trans. Knowledge Discovery from Data}, \textbf{1}, 1 (2007)\\
\noindent [74] M. Ley, in {\em Proc. Int. Symposium on String Processing and Information Retrieval} (2002), 1-10\\
\noindent [75] J. S. Coleman, {\em Introduction to mathematical sociology}, London Free Press Glencoe (1964)\\
\noindent [76] L. C. Freeman, C. M. Webster and D. M. Kirke, {\em Soc. Netw.}, \textbf{20}, 109 (1998)\\
\noindent [77] J. Coleman, E. Katz, and H. Menzel, {\em Sociometry}, \textbf{253} (1957)\\
\noindent [78] L. A. Adamic and N. Glance, in {\em Proc. Int. Workshop on Link Discov.}, 36-43, (2005)\\
\noindent [79] M. D. Choudhury, H. Sundaram, A. John, and D. D. Seligmann, in {\em Proc. Int. Conf. on Computational Science and Engineering}, 151-158, (2009)\\
\noindent [80] M. Richardson, R. Agrawal, and P. Domingos, {\em The Semantic Web-ISWC 2003} (Springer, 2003), 351-368\\
}
\bibliographystyle{Science}
\vspace*{.2cm}
\textbf{Acknowledgments:} Particular gratitude goes to D. Fanelli and P. K. Maini for useful discussions. \textbf{Funding:} The work of M.A. and T.C. presents research results of the Belgian Network DYSCO, funded by the Interuniversity Attraction Poles Programme, initiated by the Belgian State, Science Policy Office. The work of M.A. is also supported by a FRS-FNRS Postdoctoral Fellowship.  {\textbf{Author contributions:} M.A. and T.C. designed the study, conducted the formal analysis and analysed the data. All the authors organised and wrote the manuscript. \textbf{Conflict of interest:} The authors declare that they have no competing interests. \textbf{Data and materials availability:} All data needed to evaluate the conclusions in the paper are present in the paper and/or the Supplementary Materials. Additional data related to this paper may be requested from the authors.}

\end{document}